%% file: TMM_technote.tex
\documentclass[10pt,twocolumn]{IEEEtran}

\IEEEoverridecommandlockouts                              
\let\labelindent\relax

\usepackage{amsmath} 
\usepackage{amssymb}  
\usepackage{amsfonts}
\usepackage{graphicx}
\usepackage{subfigure}
\usepackage[usenames]{color}
\usepackage[shortlabels]{enumitem}

\title{\LARGE \bf
Convex Parameterizations and Fidelity Bounds for\\  Nonlinear Identification and Reduced-Order Modelling
}

\IEEEoverridecommandlockouts
\author{
Mark M. Tobenkin \ \ \ Ian R. Manchester \ \ \ Alexandre Megretski
\thanks{M. Tobenkin and A. Megretski are with the Department of Electrical Engineering   and   Computer   Science,   Massachusetts   Institute   of   Technology, Cambridge, MA, 02139. Email: \{mmt, ameg\}@mit.edu} \thanks{I. Manchester is with the Australian Centre for Field Robotics and  School of Aerospace, Mechanical and Mechatronic Engineering, University of Sydney, NSW 2006, Australia. Email: ian.manchester@sydney.edu.au.}
\thanks{This was supported by National Science Foundation Grant
  No. 0835947, the LANL Information Science and Technology Institute, and the Australian Research Council (DP130100551). }
}

\newtheorem{prob}{Problem}
\newtheorem{thm}{Theorem}

\newtheorem{definition}{Definition}

\newtheorem{remark}{Remark}

\newenvironment{pf}{\smallbreak\noindent{\it Proof. }}{\hfill$\Box$\smallbreak}
\newenvironment{pf*}[1]{\smallbreak\noindent{\it #1}}{\hfill$\Box$\smallbreak}

\makeatother
\ifx\Box\undefined
  \makeatletter\input{oldlfont.sty} \makeatother
\fi

\newcommand{\cl}[1]{{\cal #1}}
\newcommand{\pd}[2]{\frac{\partial #1}{\partial #2}}

\def\Box{\hbox{\hskip 1pt \vrule width 8pt height 6pt depth 1.5pt
  \hskip 1pt}}

\newcommand{\e}{\epsilon}

\newcommand{\D}{\Delta}

\renewcommand{\r}{\rho}

\newcommand{ \R}{\mathbb R}

\newcommand{\Pinv}{{(P^{-1})}}
\newcommand{\M}{\mathcal M}

\newcommand{\EQ}[2]{\begin{equation}\label{#1}#2\end{equation}}

\newcommand{\RR}{\mathbb{R}}

\usepackage{verbatim}

\begin{document}

\maketitle
\thispagestyle{empty}
\pagestyle{empty}

\begin{abstract}
\input{tacAbstract}
\end{abstract}

\section{Introduction}
\input{tacIntroShort}

\section{Problem Setup \& Preliminaries}
\label{sec:setup}
\input{tacSetup}
\input{tacLagrangian}

%

\section{Convex Parameterization of Stable Models}
\label{sec:modelclass}
\input{tacModelClass}
\section{Convex Upper Bounds for Simulation Error}
\label{sec:upperbounds}
\input{tacUpperBounds}
\section{Contracting Models and Linearized Simulation Error}
\label{sec:linearized}
\input{tacLinearized}
 \section{Examples and Discussion}
 \label{sec:examples}
 \input{tacExamples}

 \section{Acknowledgments}
 \input{cdc2010MIack}

\bibliographystyle{IEEEtran}
\bibliography{elib}

\end{document}

%% file: tacAbstract.tex
Model instability and poor prediction of long-term behavior are common
problems when modeling dynamical systems using nonlinear ``black-box''
techniques. Direct optimization of the
long-term predictions, often called \emph{simulation error
minimization}, leads to optimization problems that are generally non-convex in the
model parameters and suffer from multiple local minima. In this work we present methods which address
these problems through convex optimization, based on Lagrangian relaxation, dissipation inequalities, contraction theory, and  semidefinite programming.  We demonstrate the proposed methods with a model
order reduction task for electronic circuit design and the
identification of a pneumatic actuator from experiment.

%% file: tacIntroShort.tex
Building approximate models of dynamical systems from data is a ubiquitous task in the sciences and engineering. Reduced order modelling and system identification are two well-known cases, which differ mainly in the source of the data. In reduced order modelling the data come from simulation ``snapshots'' of a complex model\footnote{We note that another class of model reduction algorithms work directly from the defining equations of the high-order model rather than simulation snapshots, e.g. balanced truncation and Hankel-optimal reduction \cite{antoulas2005approximation}.}, which is usually based on physical principles (e.g. \cite{Holmes98, Lall02, lucia2004reduced}). The central problem is finding a model that simulates accurately despite unmodelled dynamics. In system identification the data come from experimental recordings \cite{Sjoberg95, Ljung99, pearson2006nonlinear, Ljung10}, and this brings the additional problems of measurement noise and exogenous disturbances. When some known structure can be incorporated into a model it is usually preferable to do so, but black-box modeling
plays an important role when first-principles models are either
weakly identifiable, too complicated for the eventual application, or
simply unavailable. 


The basic objective we consider is to obtain a dynamical system that accurately reproduces the observed relationship between some given input and output data, and can reliably generalize to other inputs.  A complete solution using optimization methods requires a specification of a model class, a choice of objective function, and ultimately an algorithm for minimizing the objective function over the model class.

The model class we propose consists of discrete-time nonlinear state space models. Such models have previously been studied in the context of model reduction (e.g. \cite{Lall02}) and system identification (e.g. \cite{Schon10}). It includes all linear time-invariant (LTI) systems, systems defined by LTI internal dynamics with nonlinear input/output mappings (e.g. Wiener and Hammerstein models \cite{billings1982identification}), and includes nonlinear autoregressive models (e.g. \cite{Sjoberg95, billings2013nonlinear}) by choosing the state to be a truncated history of outputs.


An alternative is to model the current output as a static function of a finite history of inputs, e.g. Nonlinear Finite-Impulse-Response (NFIR) and Volterra models \cite{Sjoberg95, Doyle02}. However, such models are inefficient at representing inherently dynamic behaviour such as resonance, which is naturally expressed by models with internal feedback.


The inclusion of feedback in a model means that stability is no longer guaranteed, and in many applications an unstable model would be unacceptable. 
The ability of a model to ``generalize'' to inputs not in the training data depends critically on model stability. 
Ensuring stability of models with feedback is a significant problem and has been studied for many years for linear system identification \cite{Ljung10}.
 It is known that stability of linear autoregressive models can   be guaranteed if the input is generated by an autoregressive process and the data sequence is sufficiently long \cite{Regalia95}. 
Stability of linear state space models generated by subspace identification was addressed by padding a shifted state matrix with zeros in \cite{maciejowski1995guaranteed}, and by regularization in \cite{van2001identification}. A linear matrix inequality (LMI) parametrization of stable linear models was given in \cite{Lacy03}.  

The challenge of ensuring stability of nonlinear models with feedback is discussed in \cite{pearson2006nonlinear,Ljung10,billings2013nonlinear}. To the authors' knowledge, there are few prior published methods that guarantee stable models regardless of the source of the data. Worthy of note are the recent results of \cite{besselink2013model} giving small-gain and passivity-like stability guarantees for reduction of a high-order linear system in feedback with a convergent nonlinear system.



The objective function we focus on in this paper is simulation error, a.k.a. output error. This function is in general highly non-convex for models with feedback, and it is challenging to compute the global minimum \cite{Ljung10}.  Indeed, one can find poor local minima for the class of second-order linear models  \cite{soderstrom1975uniqueness}.

A computationally simpler alternative is to minimize the {\em equation error}, i.e. the one-step ahead prediction error, rather than the long-term simulation error.
%
Nonlinear ARX and subspace identification, for example, use these criteria \cite{Sjoberg95, Ljung99}. In many cases, the minimization amounts to basic least squares and can be solved very efficiently.  The disadvantage is that there is in general no connection between the equation error and the simulation error, since an unstable (or weakly stable) model might be found for which small equation errors accumulate under recursive simulation.

The main contributions of this paper are a class of nonlinear state-space models that is convex in its parameters and guarantees stability, and a family of convex upper bounds for minimization of simulation error. The combination of these allows problems in reduced order modelling and identification to be approached using semidefinite programming. Our approach builds upon the ideas proposed in \cite{Megretski08} for nonlinear autoregressive models, which was extended and applied to electronics model reduction in \cite{Bond10}. The model set and some fidelity bounds in this paper were reported without proof in \cite{tobenkin2010convex}. The present paper significantly extends these results by introducing more accurate fidelity bounds based on Lagrangian relaxation, complete proofs of technical results, and new experimental results. We do not discuss the effects of measurement noise, but we note that a modified version of the proposed method gives consistent estimates when repeated measurements are available \cite{manchester2012stable, tobenkin2013stable,tobenkin2014robustness}.

%% file: tacSetup.tex

Spaces of vector signals, i.e. mappings $\{0, 1, ..., T\}\rightarrow \RR^n$ are denoted $l_T^n$. For a symmetric matrix $A=A'$,  $A \ge 0$ and  $A>0$ mean that $A$ is positive semidefinite and positive-definite, respectively. Inequalities $A\ge B, A>B$ mean $A-B\ge 0, A-B>0$, respectively. For symmetric positive-semidefinite $Q$ the notation $|v|_Q^2$ is shorthand for $v'Qv$. 
Throughout the paper, signals with a tilde, e.g. $\tilde y(t)$, refer to recorded data, whereas signals without, e.g. $y(t)$ refer to the corresponding quantity generated by an identified model. 


Assume that training data are given in the form of finite sequences of inputs and outputs: $\tilde u = \{\tilde u(0), \tilde u(1), ..., \tilde u(T)\}\in l_T^m$, $\tilde y = \{\tilde y(0), \tilde y(1), ..., \tilde y(T)\}\in l_T^p$. We use the notation $\tilde z: = \{\tilde u, \tilde y\}$ to represent the ``available data''. 

In this paper, a model class is a triple $\M = (a, g, \Theta)$ where $a:\RR^n\times\RR^m \times\RR^q\rightarrow \RR^n$ and $g:\RR^n\times\RR^m \times\RR^q\rightarrow \RR^p$ are continuously differentiable functions, and $\Theta \subset \R^q$ is a set of $q$-dimensional parameter vectors. A model class $(a, g, \Theta)$ is associated with state-space dynamical systems of the form
\begin{eqnarray}
x(t+1)&=&a(x(t),u(t),\theta),\label{exsys1}\\
y(t)&=&g(x(t),u(t), \theta),\label{exsys2}
\end{eqnarray}
where $x\in\RR^n, u\in\RR^m, y\in\RR^p$, $\theta\in\Theta$.
%
It is desirable that identified models generalize to inputs not in the training data, it is appropriate to impose a strong notion of stability that which ensures predictable response to a wide variety of inputs independent of model initial conditions, such as the following:
\begin{definition}\label{def:incL2} A system \eqref{exsys1}, \eqref{exsys2} is {\em incrementally $\ell^2$ stable} if for every pair of initial conditions $x_1(0), x_2(0)$ and every input sequence $u(t), t = 1, 2, ...$, the states $x_1(t), x_2(t)$ and outputs $y_1(t) = g(x_1(t), u(t))$ and $y_2(t) = g(x_2(t), u(t))$ satisfy
$
\sum_{t=0}^{\infty}|x_1(t)-x_2(t)|^2 < \infty, \ \ \sum_{t=0}^{\infty}|y_1(t)-y_2(t)|^2 < \infty.
$
\end{definition}

In addition to a model set, one must define a measure of model fidelity. There are many possibilities, but perhaps the most natural and straightforward measure of quality is {\em simulation error}, also known as {\em output error}:
\begin{equation}\label{eqn:J_se}
J_{se}(\theta,x,\tilde z) := \sum_{t=0}^T |y(t)-\tilde y(t)|^2
\end{equation}
which is to be minimized over parameters $\theta$ and states $x$ for given data $\tilde z$, subject to dynamic constraints \eqref{exsys1}, \eqref{exsys2}. 

The major difficulties we seek to address in this paper are:
\begin{enumerate}[(a)]
\item Obtaining a useful set of stable models. For all non-trivial parameterizations of functions $a(x(t),u(t),\theta)$, an explicit and tractable characterization of the set of all $\theta$ defining a stable model is unknown. 
\item Efficient global minimization of $J_{se}$. Even if the cost function $J(\theta,x,\tilde z)$ is convex in $\theta$, the dynamic constraints \eqref{exsys1} generally define a nonconvex set: the simulation output is generated by the recursion
\[
y(t) = g(a(a(... a(x(0),\tilde u(0), \theta),..., \tilde u(t-1), \theta) , \tilde u(t),\theta),
\]
so even if $a$ and $g$ are linear in the parameter $\theta$, $J_{se}$ is a highly nonlinear function of the model 
\end{enumerate}
In this paper, we construct a family of computationally tractable approximations to the simulation error minimization problem. More specifically, we formulate model classes $\hat {\mathcal M} = (a, g, \hat \Theta)$ and optimization problems of the form:
$\min_{\theta\in\hat\Theta}  \hat J(\theta, \tilde z),$ 
with the following properties:
\begin{enumerate}[(a)]
\item $\hat\M $ is a finite-dimensional  convex parametrization of models satisfying the following inclusions:
$
LTI \subseteq \hat\M \subseteq \M_{\ell 2}
$
where $LTI$ refers to the set of all stable linear time invariant models of appropriate dimension, and $\M_{\ell 2}$ is the set of all (possibly nonlinear) models that are {\em incrementally $\ell^2$ stable}.
\item $\hat J$ is a convex function of $\theta$ such that: (i) $
 \hat J(\theta, \tilde z)\ge J_{se}(\theta, x, \tilde z)
$ for any feasible $x$; (ii) $\hat J(\theta, \tilde z)$ always has a finite minimum over $\theta$; (iii) $\hat J(\theta, \tilde z)=0$ when $\tilde z$ are generated by simulating the model represented by $\theta$. In other words, the method perfectly recovers a ``true'' model if one exists in the model class.
\end{enumerate}





%% file: tacLagrangian.tex

A key tool we use is the Lagrangian relaxation, or S-Procedure, which is a method for approximating optimization problems with ``difficult'' nonconvex constraints by ``easier'' unconstrained problems with multipliers. Versions of this method have been used extensively in robust control and combinatorial optimization \cite{polik2007survey, lemarechal2001lagrangian}. In the context of this paper the idea is to approximate problems of the form
\begin{align}
J^\star = \min_{x, \theta} J(\theta, x), \textrm{ subject to } h(\theta, x) = 0
\end{align}
where $h(\theta, x)$ is a vector of constraint functions. In the context of simulation-error minimization, $J(\theta, x)$ represents the measure of model fit, e.g. \eqref{eqn:J_se}, and $h(\theta, x)$ represents the dynamical constraints, e.g. \eqref{exsys1}. 

Suppose that $J$ is a convex function of $\theta$ and $h$ is an affine function of $\theta$, but both are nonconvex functions of $\theta$ and $x$ jointly. In our context, Lagrangian relaxation refers to the approximation of the above problem by the following:
\begin{equation}\label{eq:lagrange}
\hat J_\lambda^\star= \min_\theta \hat J_\lambda(\theta), \quad \hat J_\lambda(\theta) :=\sup_x\{J(\theta, x)+\lambda(x)'h(\theta, x)\}
\end{equation}
where $\lambda$ is some vector function analogous to a Lagrange multiplier in nonlinear programming. The function $\hat J_\lambda(\theta)$ is clearly convex, since it is the supremum of a family of convex functions \cite{boyd2009convex}. It is also an upper bound for $J(\theta,x)$ for any feasible $x$: if $h(\theta,x)=0$  in \eqref{eq:lagrange}, then $\hat J_\lambda(\theta) =J(\theta,x)$, so the supremum over $x$ must be no smaller.


It is apparent that if the Lagrangian relaxation \eqref{eq:lagrange} is applied using two different constraint functions that nevertheless have the same feasible set: $h_1(\theta, x)=0 \Leftrightarrow h_2(\theta, x)=0$, but which differ for other values of $\theta, x$, then the resulting values of $\hat J_\lambda(\theta)$ can differ. Consequently, performance of the Lagrangian relaxation may be improved by allowing a {\em redundant} parametrization of dynamical systems. 

We will frequently deal with convex functions of the form $f(c, P) = c'P^{-1}c$, where $c$ and $P$ are decision variables, $c$ is a vector and $P$ is a symmetric positive-definite matrix of appropriate size. These functions can be represented by a slack variable and an LMI using the Schur complement \cite{boyd2009convex}. Concave functions $g(c,P)=-c'P^{-1}c$ with $P>0$ obey the following upper bound:
\begin{equation}\label{eq:quadbound}
-c'P^{-1}c\le b'Pb-2b'c
\end{equation}
where the right-hand side is a convex function of $c, P$ for any fixed $b$. Inequality \eqref{eq:quadbound} 
follows directly from the expansion
$
 b'Pb-2b'c+c'P^{-1}c =   |c-Pb|^2_{P^{-1}} \ge 0.
$
From this expansion it is also clear that the bound  \eqref{eq:quadbound}  is tight if $c=Pb$.

%% file: tacModelClass.tex

The models we propose are constructed from three linearly parametrized vector functions:
\begin{align} \notag
&e_\theta(x) = \sum_{i = 0}^q \theta_i e_i(x), \ \ \ \ \
f_\theta(x,u) = \sum_{i = 0}^q \theta_i f_i(x,u), \\
& g_\theta(x,u) = \sum_{i = 0}^q \theta_i g_i(x,u),\label{eq:linparam}
\end{align}
where $\theta\in\RR^q$ and each basis function $e_i:\ \RR^n\to\RR^n, f_i:\ \RR^n\times\RR^m\to\RR^n, g_i:\ \RR^n\times\RR^m\to\RR^p$ is continuously differentiable in its arguments.
These functions are associated with a parametrization of implicit state space dynamic models
\begin{eqnarray}
e_\theta(x(t+1))&=&f_\theta(x(t),u(t)),\label{mi5}\\
y(t)&=&g_\theta(x(t),u(t)).\label{mi6}
\end{eqnarray}
A particular choice of state dimension
and bases for $e, f, g$ defines a parametrization of models of the form \eqref{exsys1}, \eqref{exsys2}  given by $a(x,u,\theta) = e_\theta^{-1}(f_\theta(x,u))$ and $g(x,u,\theta) = g_\theta(x,u)$, as long as it can be guaranteed that $e_\theta(x)$ is a bijection for all $\theta$.


We denote by $\mathcal E, \mathcal F, \mathcal G$ the span of the functions \eqref{eq:linparam} over $\theta \in \RR^q$. We assume that $\mathcal E, \mathcal F, \mathcal G$ contain all linear functions $x \mapsto Ex, (x,u)\mapsto Ax+Bu, (x,u) \mapsto Cx+Du$, respectively,  where $A, B, C, D, E$ are matrices of appropriate dimensions, and that $\mathcal F$ is closed under multiplication by a non-singular $n\times n$ matrix, i.e. if $f(\cdot,\cdot)\in \mathcal F$ then $Mf(\cdot,\cdot)\in\ F$ for all $M\in\RR^{n\times n}$ with $\det(M)\ne 0$. Simple choices for $\mathcal E, \mathcal F, \mathcal G$ satisfying these assumptions are vectors of multivariate polynomials or trigonometric polynomials up to some fixed degree, parametrized by their coefficients. For the sake of brevity, we will usually drop the subscript $\theta$ from the functions $e, f$ and $g$ and speak of searching directly for $e, f$ and $g$.

To ensure model stability, we propose the following constraint, requiring an auxiliary matrix variable $P=P'>0$:
\begin{align}
& |f(x_1,u)-f(x_2,u)|^2_{\Pinv } -2(x_1-x_2)'(e(x_1)-e(x_2)) \notag\\ &+|x_1-x_2|_{P+\mu I}^2+|g(x_1, u)-g(x_2,u)|^2 \le 0 \label{eq:stab}
\end{align}
for all $x_1, x_2 \in \RR^n, u \in \RR^m$ for some fixed $\mu>0$. Note that this constraint is convex with respect to $e, f, g, P$ and hence convex with respect to a parameter vector $\theta$ when $e, f, g, P$ are affinely parametrized. 
\begin{definition}\label{def:modelclass} For a given choice of state dimension and basis functions $e_i, f_i, g_i$, let $\Theta$ be the set of $\theta\in\RR^q$ for which the model \eqref{mi5}, \eqref{mi6} satisfies the inequality \eqref{eq:stab} for some $P=P'>0$. Additionally, let $\hat\M$ be the model class $(e_\theta^{-1}(f_\theta(x,u)), g_\theta (x,u), \Theta)$.
\end{definition}


Model well-posedness, i.e. existence of a unique solution,
is guaranteed by the following theorem:
\begin{thm}\label{thm:wellpos}
Given a model \eqref{mi5}, \eqref{mi6}, if
\eqref{eq:stab} holds for all $x_1, x_2, u$, then
$e$
is a bijection $\RR^n \leftrightarrow \RR^n$.  

\end{thm}
\begin{pf}
We need to show that for every $z\in\R^n$, the equation $e(x)=z$ has a unique solution. Firstly, note that \eqref{eq:stab} implies 
$
2(x_1-x_2)'(e(x_1)-e(x_2)) \ge \mu|x_1-x_2|^2.
$
It is known (see, e.g., \cite[Th. 18.15]{poznyak2010advanced}) that if $T:\R^n\to\R^n$ is a nonlinear operator satisfying $(x_1-x_2)'(T(x_1)-T(x_2))\ge c|x_1-x_2|^2$ for all $x_1, x_2\in\R^n$ with $c>0$, then the nonlinear equation $T(x)=0$ has a unique solution. Take $T(x) = e(x)-z$ and $c = \mu/2$ and the theorem is proved.
%
\end{pf}

For simulation of the model, a solution to the equation $e(x)=z$ can be found by application of the Newton method. Convergence can be guaranteed by guarding with the ellipsoid method
\cite[Sec 2.2.2]{tobenkin2014robustness}.

\begin{thm}\label{thm:stability}
For any model of the form  \eqref{mi5}, \eqref{mi6}, the condition  \eqref{eq:stab} implies global incremental $\ell^2$ stability.  
\end{thm}
\begin{pf}
Consider the incremental Lyapunov function candidate $V(x_1, x_2) = |e(x_1)-e(x_2)|_{\Pinv}^2$ with $P=P'>0$. The inequality
\begin{align}
-|e(x_1)-e(x_2)|^2_\Pinv +|f(x_1,u)-f(x_2,u)|^2_\Pinv &\notag\\ +\mu|x_1-x_2|^2+|g(x_1, u)-g(x_2,u)|^2 &\le 0 \label{eq:lyap_nonrelaxed}
\end{align}
for all  $x_1, x_2\in \RR^n, u\in\RR^m$ is a sufficient condition for global incremental $\ell^2$ stability.
Now apply the bound \eqref{eq:quadbound}, with $c = e(x_1)-e(x_2)$ and $b = x_1-x_2$, giving
\[
-|e(x_1)-e(x_2)|^2_\Pinv \le |x_1-x_2|_{P}^2-2(x_1-x_2)'(e(x_1)-e(x_2)).
\]
and substitute the right-hand-side of this inequality for $-|e(x_1)-e(x_2)|^2_\Pinv$ in \eqref{eq:lyap_nonrelaxed}.
It is then clear that \eqref{eq:stab} implies \eqref{eq:lyap_nonrelaxed} and hence stability of the model.
\end{pf}

\begin{remark} It is straightforward to impose other behavioral constraints, e.g. passivity, by the same convexification procedure and appropriate choice of supply rate. 
\end{remark}


\begin{definition} A system 
$
  x(t+1) =  a(x(t),u(t)),
  y(t) = g(x(t),u(t)), $
defined by functions 
 $a: \RR^n \times \RR^m \mapsto \RR^n$ and $g:  \RR^n \times \RR^m \mapsto \RR^p$ is called
quadratically incrementally $\ell^2$ stable, or more briefly: quadratically stable, if there exists an $M \in \RR^{n \times n}$ with $M = M' > 0$ 
such that the condition
\begin{equation}
 |a(x_1,u) - a(x_2,u)|^2_M - |x_1-x_2|_M^2 \leq - |g(x_1,u)-g(x_2,u)|^2,
  \label{eq:quadstab}
\end{equation}
holds for all $x_1, x_2 \in \RR^n$ and $u \in \RR^m$. I.e. incremental stability can be verified by a quadratic Lyapunov function.

\end{definition}


\begin{thm}\label{lem:coverage}
The model set $\hat {\mathcal M}$ includes all quadratically incrementally $\ell^2$ stable models of the form \eqref{exsys1}, \eqref{exsys2} for which $a(\cdot, \cdot)\in \mathcal F$ and $g(\cdot, \cdot)\in\mathcal G$.
\end{thm}
\begin{pf}
Suppose \eqref{eq:quadstab} holds for some $M=M'>0$.
Choose $e(x) = Mx, f(x,u) = Ma(x,u)$, and $P = M$, i.e. $c=Pb$ in  \eqref{eq:quadbound}, and leave $g(x,u)$ as is. Clearly this model is equivalent to \eqref{exsys1}, \eqref{exsys2}, and by assumption that $a(\cdot,\cdot) \in \mathcal F$ and that $\mathcal F$ is closed under multiplication by $M$, $f(\cdot,\cdot)\in \mathcal F$. Direct substitution into \eqref{eq:stab} reveals that if \eqref{eq:quadstab} holds then so does \eqref{eq:stab}.
\end{pf}

\begin{remark}
	All stable linear systems are automatically stable in the above sense, as are models with stable linear dynamics but nonlinear input and/or output maps, e.g.  Wiener and Hammerstein models, as long as $g(x,u)$ is globally Lipschitz in $x$.
\end{remark}


%


%% file: tacUpperBounds.tex
Let us suppose there is method for generating a vector sequence $\tilde x\in\ell_T^n$ from available data $\tilde z$. For our main theoretical results $\tilde x$ can be completely arbitrary, but it will be shown below in Theorem \ref{thm:riegdt} that best results are obtained if $\tilde x$ approximates a good state sequence for the model, hence we refer to $\tilde x$ as the {\em surrogate state} sequence. Typical methods for generating $\tilde x$ include subspace identification \cite{van1996subspace}, using a history of outputs as in nonlinear ARX \cite{billings1982identification}, and -- in the context of model reduction -- proper orthogonal decomposition \cite{Holmes98}. With some abuse of notation, from now on we consider the available data set to include $\tilde x$, i.e. $\tilde z := \{\tilde u, \tilde y, \tilde x\}$.

Given a surrogate state sequence, we will frequently refer to the following {\em equation error} functions:
\begin{align}
\epsilon(\theta, \tilde z, t) &= e_\theta(\tilde x(t+1))-f_\theta(\tilde x(t), \tilde u(t)), \label{eqn:eps_x}\\
\eta(\theta, \tilde z, t) & = \tilde y(t) - g_\theta(\tilde x(t), \tilde u(t)), \label{eqn:eps_y}
\end{align}
although for brevity of notation we suppress the arguments $(\theta, \tilde z, t)$ and simply denote dependence on time by subscripts: $\epsilon_t, \eta_t$, and similarly for other signals e.g. $x_t=x(t)$.

Equation error $\epsilon_t$ is the one-step prediction error in a coordinate system generated by the bijection $e(\cdot)$. 
A straightforward way to search for a model is to minimize the equation error in a least-squares sense:
$
J_{EE}(\theta,\tilde z) = \sum_{t=0}^T |\epsilon_t|^2+|\eta_t|^2,
$
which is clearly convex if models are affinely parameterized.
E.g. \cite{Lacy03} suggests identifying linear models in precisely this way. However, while this is often effective there is in general no direct relationship between $J_{EE}$ and the (open-loop) simulation error, which results from recursion as described in Section \ref{sec:setup}.

Application of the Lagrangian relaxation \eqref{eq:lagrange} suggests the following convex upper bound for simulation error, which is valid for any choice of $n$-dimensional vector function $\lambda(x,\tilde z, t)$:
\begin{align}
\sup_{x\in l_T^n}\bigg\{&\sum_{t=0}^{T}|g(x_t, \tilde u_t)-\tilde y_t|^2  +\sum_{t=0}^{T-1}\lambda'\big(e(x_{t+1})-f(x_t,\tilde u_t)\big)\bigg\}.\notag
\end{align}

Our first proposal for a convex upper bound on simulation error comes from choosing $\lambda(x,\tilde z, t) = -2(x_{t+1}-\tilde x_{t+1})$:
\begin{align}
\hat J_L(\theta, \tilde z):= &\sup_{x\in l_T^n} \bigg\{\sum_{t=0}^{T}|g(x_t, \tilde u_t)-\tilde y_t|^2 \notag \\  & -\sum_{t=0}^{T-1}2(x_{t+1}-\tilde x_{t+1})'(e(x_{t+1})-f(x_t,\tilde u_t)) \bigg\}, \notag \\ & \textrm{ subject to }x(0) = \tilde x(0). \label{eq:Lupper}
\end{align}
With this upper bound and the model set from Definition \ref{def:modelclass}, we construct the following convex optimization problem. 
\begin{prob}\label{prob:globalMultistep} For a model class $(a, g, \Theta)$ as in Definition \ref{def:modelclass} and data set $\tilde z$, solve
$
\hat J_L^\star (\tilde z)=\min_{\theta \in \Theta} \hat J_L(\theta, \tilde z).
$
\end{prob}

A challenge in solving Problem \ref{prob:globalMultistep} is that it involves computing the supremum of a function of a large number of variables, the entire sequence $x$. 
An upper bound can be constructed by interchanging the order of supremum and summation, and introducing a positive-semidefinite storage function $V(x,t)$:
\begin{align}
J(\theta, \tilde z) \le& \sum_{t=0}^{T} \sup_{x_t, x_{t+1}}\Big\{|g(x_t, \tilde u_t)-\tilde y_t|^2 \notag \\ & +\lambda(x,\tilde z, t)'(e(x_{t+1})-f(x_t,\tilde u_t)) \notag\\ &+V(x_{t+1},t+1)-V(x_{t},t)\Big\}. \label{eq:boundLV}
\end{align}
Again, the quality of the bound will depend on the choice of functions $V$ and $\lambda$.


It will be shown below that a particular choice of $V$ and $\lambda$ leads to the upper bound
$
\hat J_V(\theta, \tilde z) := \sum_{t=0}^T\cl E(\theta, \tilde z,t),
$
where
\begin{align}\label{eq:Vupper}
\cl E(\theta, \tilde z,& t) := \sup_{x\in \RR^n} \{ |f(x, \tilde u_t)-f(\tilde x_t, \tilde u_t)-\e_{t}|_\Pinv^2 + |x-\tilde x_t|_{P}^2\notag\\ &-2(x-\tilde x_t)'(e(x)-e(\tilde x_t)) + |g(x, \tilde u_t)-\tilde y_t|^2  \} 
\end{align}
and $\e_{t}$ is given by \eqref{eqn:eps_x}. We define the following optimization problem, which is convex in $\theta$.
\begin{prob}\label{prob:globalOnestep} For a model class $(a, g, \Theta)$ as in Definition \ref{def:modelclass} and data set $\tilde z$, solve
$
\hat J_V^\star (\tilde z)=\min_{\theta\in\Theta} \hat J_V(\theta, \tilde z).
$
\end{prob}
This upper bound is the {\em robust identification error} (RIE), first introduced in \cite{tobenkin2010convex}. The following definition characterizes data sets for which ``the true system is in the model class'' \cite{Ljung10}.
\begin{definition}
Given a model class $\hat \M = (a, g, \Theta)$ and a data set $\tilde z$ we say that $\tilde z$ is in the {\em range} of $\hat \M$ if there exists a model in $\hat \M$ satisfying $\epsilon(t)=\eta(t)=0$ for all $t = 0, 1, ... T$. That is, simulating this model with the observed input and initial conditions exactly reproduces the observed state and output.
\end{definition}

We are now ready to state the main theoretical result regarding Problems \ref{prob:globalMultistep} and \ref{prob:globalOnestep}.

\begin{thm}\label{thm:riegdt}  Given a model class $\hat \M$ and a data set $\tilde z$, Let $J_{se}^\star(\tilde z)$ be the minimal simulation error over the model class $\hat \M$, then for any data set $\tilde z$ and any model parameters $\theta$ in $\hat \M$, 
$
J_{se}(\theta, \tilde z) \le \hat J_L(\theta, \tilde z) \le \hat J_V(\theta, \tilde z) 
$
and $\min_{\theta\in\hat\M} \hat J_V(\theta, \tilde z) <\infty$.
Furthermore, if the data set $\tilde z$ is in the range of $\hat \M$, then 
$
J_{se}^\star(\tilde z) = \hat J_L^\star(\tilde z) = \hat J_V^\star(\tilde z) =0.
$
\end{thm}
\begin{pf}
The statement that $J_{se}(\theta, \tilde z) \le  \hat J_L(\theta, \tilde z)$ for all $\theta$ follows from the basic properties of the Lagrangian relaxation.

To analyze $\hat J_V(\theta, \tilde z)$ we first show that it corresponds to \eqref{eq:boundLV} with a particular choice of storage function and multiplier. Indeed, let $V(x,t) = 2(x-\tilde x_t)'(e(x)-e(\tilde x_t)) -|x-\tilde x_t|_{P}$ and $\lambda_t = -2(x_{t+1}-\tilde x_{t+1})$, then \eqref{eq:boundLV} reduces to:
\begin{align}
&\sup_{x_t, x_{t+1}}\{|g(x_t, \tilde u_t)-\tilde y_t|^2 -|x_{t+1}-\tilde x_{t+1}|_{P}  \notag\\ &+2(x_{t+1}-\tilde x_{t+1})'(f(x_t, \tilde u_t) -f(\tilde x_t, \tilde u_t)-\e_{t})  \notag\\ &-2(x_t-\tilde x_t)'(e(x_t)-e(\tilde x_t)) +|x_t-\tilde x_t|_{P}\}. \label{eq:prfbound}
\end{align}
With this formulation, $x_{t+1}$ only appears in quadratic or linear terms, so the partial maximum can be computed exactly:
$
|f(x_t, \tilde u_t)-f(\tilde x_t, \tilde u_t)-\e_{t}|_\Pinv^2 = \max_{x_{t+1}}\{ -|x_{t+1}-\tilde x_{t+1}|^2_{P} \\  +2(x_{t+1}-\tilde x_{t+1})'(f(x_t, \tilde u_t)-f(\tilde x_t, \tilde u_t)-\e_{t})  \}
$
giving the following bound in place of \eqref{eq:prfbound}:
$
\sup_{x(t)}\{|g(x_t, \tilde u_t)-\tilde y_t|^2 +|f(x_t, \tilde u_t)-f(\tilde x_t, \tilde u_t)-\e_{t}|_\Pinv^2 -2(x_t-\tilde x_t)'(e(x_t)-e(\tilde x_t)) +|x_t-\tilde x_t|_{P}\}
$
which is $\cl E(\theta, \tilde z, t)$ in \eqref{eq:Vupper}.

To prove the statement $\hat J_L(\theta, \tilde z) \le \hat J_V(\theta, \tilde z)$, we first apply \eqref{eq:boundLV} with the choice of $V$ and $\lambda$ given above, to give
\begin{align}
\sup _{x}\bigg\{\sum_{t=0}^T &\big(V(x_{t+1},t+1)-V(x_t,t)+|g(x_t, \tilde u_t) - \tilde y_t|^2\notag\\  &+\lambda_t'(e(x_{t+1})-f(x_t, \tilde u_t))\big)\bigg\} \le\hat J_V(\theta, \tilde z). \label{eq:JVJL}
\end{align}
All storage function terms cancel in the summation except the first and the last. By construction\footnote{Alternatively, when working without a known initial condition, one can remove the constraint $x(0)=\tilde x(0)$ and add $-V(x_0,\tilde x_0)=-2(x_0-\tilde x_0)'(e(x_0)-e(\tilde x_0)) +|x_0-\tilde x_0|_{P}$ to terms inside the brackets in \eqref{eq:Lupper}.}, with $x(0)=\tilde x(0)$ we have $V(x(0),0)=0$, and since $V$ is positive-definite $V(x(T),T)\ge 0$, the left-hand side of \eqref{eq:JVJL} is an upper bound for 
$
\sup _{x}\bigg\{\sum_{t=0}^T \big(|g(x_t, \tilde u_t) - \tilde y_t|^2  +\lambda_t'(e(x_{t+1})-f(x_t, \tilde u_t))\big)\bigg\}
$
which equals $\hat J_L$ plus an additional multiplier term. This term clearly does not decrease the supremum, since the choice $x_{t+1} = e^{-1}(f(x_t,\tilde u_t))$ would make it zero for any choice of $x_t$.
This proves that $\hat J_L(\theta, \tilde z) \le \hat J_V(\theta, \tilde z)$.

To prove that the minimal value of $\hat J_V(\theta, \tilde z)$ is finite, note that one can choose a model \eqref{eq:linparam} with $f(x, u) = 0, g(x,u)=0$. In this case, the bound \eqref{eq:Vupper} reduces to
$
\cl E(\theta, \tilde z, t) = \sup_{x\in \RR^n} \{|\tilde y_t|^2 +|\e_{t}|_\Pinv^2 + |x-\tilde x_t|_{P}^2 -2(x-\tilde x_t)'(e(x)-e(\tilde x_t)) \}
$
and  the stability constraint \eqref{eq:stab} ensures that 
$
|x-\tilde x_t|_{P}^2 -2(x-\tilde x_t)'(e(x)-e(\tilde x_t))\le-\mu |x - \tilde x_t|^2
$
for all $x, t$. Since $\tilde y_t$ and $\e_{t}$ are not functions of $x$ this proves the supremum is finite.

Similarly, if $\tilde z$ is in the range of $\hat \M$, then there exists a model $\bar \theta \in\hat \M$ for which $\epsilon(t)=0, \eta(t)=0$ for all $t$. Let $\bar e, \bar f, \bar g$ be the model functions given by $\bar\theta$, then \eqref{eq:Vupper} is:
\begin{align*}
&\cl E(\bar \theta, \tilde z) = \sup_{x\in \RR^n} \{ |\bar g(x, \tilde u_t)-\bar g(\tilde x_t, \tilde u_t)|^2 + |x-\tilde x_t|_{P}^2\\ & +|\bar f(x, \tilde u_t)-\bar f(\tilde x_t, \tilde u_t)|_\Pinv^2 -2(x-\tilde x_t)'(\bar e(x)-\bar e(\tilde x_t))\}
\end{align*}
and by \eqref{eq:stab} this can be bounded by
$
\cl E(\bar \theta, \tilde z, t) \le \sup_{x\in \RR^n} \{ -\mu |x-\tilde x_t|^2\} =0. 
$
Therefore $\hat J_V(\tilde z, \bar\theta) = 0$. Now, since $0\le J_{se}(\theta, \tilde z) \le \hat J_L(\theta, \tilde z) \le \hat J_V(\theta, \tilde z)$ for all $\theta$, this model must give the minimal value of  $\hat J_V(\tilde z, \bar\theta)$, and furthermore  $ J(\tilde z, \bar\theta)=0$ so a perfect model is recovered. Note that we do not require existence of a {\em unique} perfect model.
\end{pf}

%% file: tacLinearized.tex
The upper bounds in the previous section, while convex, may be expensive to compute when the dependence of $e, f,$ and $g$ on $x$ is nonlinear. Contraction analysis \cite{Lohmiller98} provides an alternative set of tools for studying model behaviour, based on infinitesimal displacements of the system rather than pairs of solutions. The principal advantage in the context of this paper is that it results in significantly easier optimization problems involving suprema of {\em quadratic} functions, which can be computed explicitly or represented using LMIs.

\subsection{Contraction Condition for Incremental $\ell^2$ Stability}

In this section we derive contraction conditions for incremental $\ell^2$ stability. Contraction analysis is based on the study of an extended system consisting of  \eqref{mi5}, \eqref{mi6} as well as the {\em differential} dynamics
\begin{align}
E(x(t+1))\D(t+1) = F(x(t), u(t))\D(t), \label{eq:diffdyn1} \\ 
\D_y(t) = G(x(t),u(t))\D(t) \label{eq:diffdyn2},
\end{align}
i.e. the dynamics \eqref{mi5}, \eqref{mi6} linearized with respect to initial conditions, 
where $E(x) = \frac{\partial}{\partial x} e(x)$, $F(x, u) = \frac{\partial}{\partial x} f(x,u)$, and $G(x, u) = \frac{\partial}{\partial x} g(x,u)$. 

For our purposes, a differential storage function can be any positive-semidefinite function of $x$ and $\Delta$, i.e. $V:\R^n\times\RR^n\to \R$ such that $V(x,\Delta)\ge 0$ for all $x$ and $\Delta$. 
If a differential storage function can be found that verifies the ``differential'' dissipation inequality
\begin{align}
	& V(x(t+1),\Delta(t+1))-V(x(t),\Delta(t))\le \notag \\& -|G(x(t),u(t))\D_t|^2-\mu|\D(t)|^2\label{eq:diff_diss}
\end{align}
for solutions of \eqref{mi5}, \eqref{eq:diffdyn1}, then $V(x(0),\D(0))$ is an upper bound for 
$\sum_{t=0}^T|G(x(t),u(t))\D_t|^2+\mu\sum_{t=0}^T|\D(t)|^2$ for all $T$, for all solutions of the extended system \eqref{mi5}, \eqref{mi6}, \eqref{eq:diffdyn1}, \eqref{eq:diffdyn2}.

We propose to use the Riemannian metric $V(x,\D) = |E(x)\D|_\Pinv^2$ with $P=P'>0$ as a differential storage function, with which \eqref{eq:diff_diss} becomes\begin{equation}\label{eq:contraction}
|F(x,u)\D|_\Pinv^2 - |E(x)\D|_\Pinv^2 + |G(x,u)\D|^2\le -\mu |\D|^2.
\end{equation}
Just as \eqref{eq:lyap_nonrelaxed} is implied by \eqref{eq:stab}, \eqref{eq:contraction} is implied by
\begin{align}
F(x, u)'P^{-1} F(x,u) +P&\notag\\- E(x)-E(x)' + G(x, u)'G(x,u) &\le -\mu I. \label{eq:contractionConvex}
\end{align}
Furthermore, unlike  \eqref{eq:contraction}, condition \eqref{eq:contractionConvex} is jointly convex in $e,f,g$ and $P$, i.e. convex in the parameter vector $\theta$. 
We now give an alternative model class to $\hat {\mathcal M}$ in Section \ref{sec:modelclass}.
\begin{definition}\label{def:contr} For a given choice of state dimension and basis functions $e_i, f_i, g_i$, we define $\Theta^0$ as the set of $\theta\in\RR^q$ for which the model \eqref{mi5}, \eqref{mi6} satisfies the inequality \eqref{eq:contractionConvex} for some $P=P'>0$. Additionally, we define the model class $\hat\M^0:=(e_\theta^{-1}(f_\theta(x,u)), g_\theta (x,u), \Theta^0)$.
\end{definition}

The main theoretical result for this model class is the following, which is analogous to Theorems \ref{thm:wellpos}, \ref{thm:stability}, and \ref{lem:coverage}.

\begin{thm}
All models in $\hat\M^0$ are well-posed and incrementally $\ell^2$ stable. Addtionally, the set $\hat\M^0$ contains all models of the form \eqref{exsys1}, \eqref{exsys2} for which \eqref{eq:diff_diss} can be verified with a constant quadratic metric $V(x,\Delta)=\Delta'M\Delta$ for some $M=M'>0$ and $a(x,u)\in \mathcal F$.
\end{thm}

\begin{pf}
To prove well-posedness, consider any pair $x_1, x_2\in \RR^n$ and define $x_\rho=(1-\rho)x_1+\rho x_2$ with $\rho\in[0, 1]$. 
Every term on the left-hand side of \eqref{eq:contractionConvex} is positive semidefinite except $- E(x)-E(x)'$, so $ E(x)+E(x)'\ge \mu I$. Now, $(x_2-x_1)'(e(x_2)-e(x_1)) = \int_0^1 (x_2-x_1)'E(x_\rho)(x_2-x_1)d\rho\ge \frac{\mu}{2}|x_2-x_1|^2$, so well-posedness is guaranteed by the same argument as in the proof of Theorem \ref{thm:wellpos}.

To prove incremental stability, consider any pair of initial conditions $x_1(0)$ and $x_2(0)$, and consider the family of solutions to \eqref{mi5} $x_\rho$, parameterized by $\rho\in[0, 1]$, having initial conditions $x_\rho(0) = (1-\rho)x_1(0)+\rho x_2(0)$ and any fixed input sequence $\bar u(t)$. Note that for $t>0$, $x_\rho(t)$ does not necessarily linearly interpolate $x_1(t)$ and $x_2(t)$.

Now, define $\Delta_\rho(t):=\frac{\partial}{\partial \rho} x_\rho(t)$. For any $T>0$ we have

$
  \sum_{t=0}^T |y_2(t)-y_1(t)|^2 =\: \sum_{t=0}^T \left | \int_0^1 G(x_\r(t))\D_\r(t) d\r \right |^2 
\leq 
\int_0^1 \sum_{t=0}^T \left | G(x_\r(t)) \D_\r(t)\right |^2d\r
$, by the Cauchy-Schwarz inequality and interchanging order of summation and integration. 

 Summing \eqref{eq:contraction} repeatedly with $x = x_\r(t)$, $u = \bar u(t)$ and $\D = \D_\r(t)$ gives
$
  \sum_{t=0}^T |y_2(t)-y_1(t)|^2  \leq  \int_0^1 \left
    |E(x_\r(0))\D_\rho(0)\right|_\Pinv^2
   -\left |E(x_\r(T+1))\D_\r(T+1)\right|_\Pinv^2\; d\r\\
  \leq   \int_0^1 \left |E(x_\r(0))(x_2(0)-x_1(0))\right|_\Pinv^2  d\r,
$
as $\D_\r(0) = x_2(0)-x_1(0)$.  As this bound holds uniformly for all $T$ we arrive at the desired result. Exactly the same reasoning can be used to bound $\sum_{t=0}^T|x_2(t)-x_1(t)|^2$.

The proof that all models with a constant quadratic metric are in $\hat\M^0$ is similar to the proof of Theorem \ref{lem:coverage}, and follows directly from the choice $e(x) = Mx, f(x,u) = Ma(x,u)$.
\end{pf}



Condition \eqref{eq:contractionConvex} is, after taking the Schur complement of $F(x,u)'P^{-1}F(x,u)$,  a {\em point-wise} linear matrix inequality, i.e. for every pair $(x,u)$ there is an LMI. For polynomial models, this can be enforced using a sum-of-squares constraint \cite{parrilo2003semidefinite}.

\subsection{Linearized Simulation Error and its Convex Bounds}
 \label{sec:linearizeddt}
We now introduce an approximation to simulation error that is easier to optimize.
Consider a ``perturbed'' version of the model equations, motivated by \eqref{eqn:eps_x}, \eqref{eqn:eps_y}:
\begin{eqnarray}
e(x_\r(t+1))&=&f(x_\r(t),u(t)) + \r\e(t),\label{mi31}\\
y_\r(t)&=&g(x_\r(t),u(t)) + \r \eta(t),\label{mi32}
\end{eqnarray}
where $\r \in [0,1]$ and $\e(t),\eta(t)$ are the equation errors.

By construction we have $y_0(t) =
y(t)$, the simulated output, and $y_1(t) = \tilde y(t)$, the
measured output. We define the linearized simulation error as
\EQ{mi34}{ J^0(\theta, \tilde z):=\lim_{\r\to1}\frac1{(1-\r)^2}
\sum_{t=0}^N|\tilde y(t)- y_\r(t)|^2}
to quantify local sensitivity of model
equations with respect to equation errors.  Applying L'Hospital's rule twice to \eqref{mi34} we
arrive at the alternative expression
$J^0(\theta, \tilde z) = \sum_{t=0}^N |G(\tilde x(t),\tilde u(t)) \tilde\D(t)
+ \eta(t)|^2,$
where $\tilde\D(t) = \left . \pd{x_\r(t)}{\r}\right|_{\r=1}$
satisfies $\tilde\D(0)=0$ and
\begin{equation}
  E(\tilde x(t+1)) \tilde\D(t+1) = F(\tilde x(t),\tilde u(t)) \tilde\D(t) +
  \e(t).\label{eq:linsimdyn}
\end{equation}

Roughly speaking, when the simulation error is small, i.e. the model fit is good, the simulation error and the linearized simulation error will be very close, and $\tilde \D(t) \approx \tilde x(t)-x(t)$. 
Note that if the model equations are affine in $x$, then the simulation error and the linearized simulation error are identical.


A family of convex upper bounds for the linearized simulation error can be constructed in a similar way to those in Section \ref{sec:upperbounds}, so we skip some details in the construction.

The analogue of Problem \ref{prob:globalMultistep} can be given in a simple form by noting that \eqref{eq:linsimdyn} can be written in ``lifted'' representation
$
H(\theta, \tilde z)\vec\D = \vec\e(\theta, \tilde z)
$
with the stacked vectors $\vec\D:=[\tilde\D(1)', \tilde\D(2)', ..., \tilde\D(T)']'$ and $\vec\e(\theta, \tilde z): = [\e(0)', \e(1)', ... \e(T)']'$ and $H(\theta, \tilde z)$ is a sparse matrix that is affine in $\theta$ and is straightforward to construct from \eqref{eq:linsimdyn}. The analogue of \eqref{eq:Lupper} can then be written as
\begin{align*}
\hat J^0_L(\theta, \tilde z):= \sup_{\vec\D\in l_T^{nT}}&\{ |\vec G(\tilde z)\vec\D +\vec\eta(\tilde z)|^2 -2\vec\D'(H(\theta, \tilde z)\vec\D-\vec\e(\tilde z)) \}\end{align*}
where $\vec G(\tilde z)$ is a block-diagonal matrix with elements $G(\tilde z(t))$ and $\vec \eta$ is a stacked vector of output equation errors $\eta(t)$. Note that both $H$ and $\vec \e$ are affine in $\theta$, so the functional $\hat J^0_L$ is convex in $\theta$. Note also that the supremum is taken of a quadratic function of $\vec\D$, which will be shown below to be concave, and can therefore be computed explicitly. 

We can also construct an analogue of Problem \ref{prob:globalOnestep} for linearized simulation error. Define the pointwise functions
\begin{align}
{\cl E}^0(\tilde z,t):=\sup_{\D\in\RR^n}& \{|F(\tilde x,\tilde u)\D+\e|_\Pinv^2
+|\D|^2_{P} \notag \\ &-2\D' E(\tilde x)\D
+|G(\tilde x,\tilde u)\D+\eta|^2 \}, \label{mi11}
\end{align}
and the following convex cost function:
$
\hat J^0_V(\theta, \tilde z) = \sum_{t=0}^N\cl E^0(\tilde z,t)
$
, which we call the {\em local robust identification error} (local RIE).

\begin{thm}\label{thm:rieldt}
For all  $P=P'>0$ and signal data $\tilde z$, for every model satisfying \eqref{eq:contractionConvex},
$ J^0(\theta, \tilde z) \le \hat J^0_L(\theta, \tilde z)\le \hat J^0_V(\theta, \tilde z)<\infty.
$
Furthermore, if the data set $\tilde z$ is in the range of $\hat \M$, then the optimal values satisfy
$
J^{0,\star}(\tilde z) = \hat J_L^{0,\star}(\tilde z) = \hat J_V^{0,\star}(\tilde z) =0.
$

\end{thm}
\begin{pf}
First we show that the bound $\hat J^0_V$ is finite. Each term ${\cl E}^0(\tilde z)$ is the supremum of quadratic functional of $\D$, so each supremum is finite if and only if the corresponding matrix 
\begin{align*}
&F(\tilde x(t),\tilde u(t))'P^{-1} F(\tilde x(t),\tilde u(t))+P\\ 
&-E(\tilde x(t))-E(\tilde x(t))'+G(\tilde x(t),\tilde u(t))'G(\tilde x(t),\tilde u(t))
\end{align*}
is negative semidefinite. In fact, the contraction condition \eqref{eq:contraction} implies that the above matrix is 
is uniformly negative definite, which implies a finite supremum achieved by a unique $\D$.

The remainder of the proof follows from the same arguments as the proof of Theorem \ref{thm:riegdt}, but with the differential storage function
$
V(\tilde x,\D) = \D'(E(\tilde x) + E(\tilde x)'-P)\D
$
and by substituting $\tilde \Delta_t$ for $\tilde x_t- x_t$, $E(\tilde x_t)\tilde \Delta_t$ for $e(\tilde x_t)-e(x_t)$ and $F(\tilde x_t)\tilde \D_t$ for $f(\tilde x_t,u_t)-f(x_t,u_t)$.
%
%
%
\end{pf}

%% file: tacExamples.tex
In this section we demonstrate the performance of the proposed methods on two real-world applications: reduced-order dynamic inversion of an op-amp and identification of a pneumatic actuator from experiments. Further examples illustrating the method can be found in \cite{tobenkin2014robustness}.

We judge the performance of a model over $T$ time samples using percent normalized simulation error:
$J_{\textrm{perf}} = 100 {\sqrt{\sum_{t=0}^{T} | \tilde y(t) - y(t)|^2
}}/{\sqrt{\sum_{t=0}^T |\tilde y(t) - \frac{1}{T+1}
  \sum_{s=0}^T \tilde y(s)|^2}}$
where $y(t)$ is the result of simulating the model from observed
initial conditions and with the known input. When minimizing the equation error, to ensure the dynamics are well-posed,
we require  $E(x) + E(x)' - 2\mu I$ to be positive semidefinite via a
sums-of-squares constraint \cite{parrilo2003semidefinite}. Software packages such as the Yalmip toolbox \cite{Yalmip} can be used to  transform polynomial optimization problems such as those in this paper to semidefinite programs. The examples in this paper took up to ten minutes to solve using the free SDP solver Sedumi \cite{sturm1999using}, though preliminary work on specialised algorithms indicates  substantial speed-ups are possible \cite{Umenberger16}.

\subsection{Low-Order Nonlinear Dynamic Inversion of an Op-Amp}

\begin{figure}
  \centering
  \subfigure[Simplified schematic of opamp.]{\includegraphics[width=0.58\columnwidth]{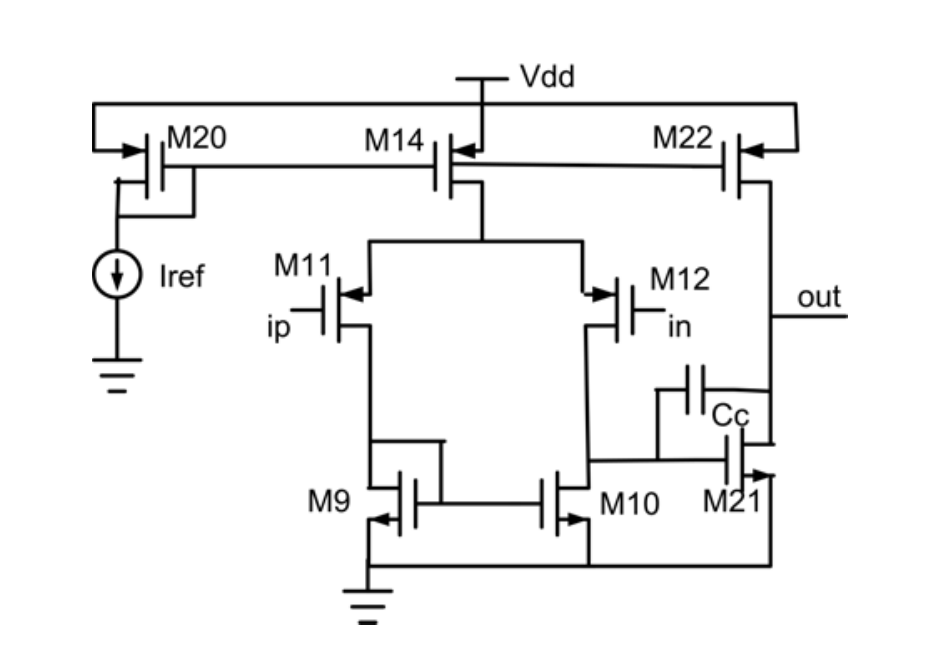}}
  \subfigure[Opamp configuration.]{\includegraphics[width=0.40\columnwidth]{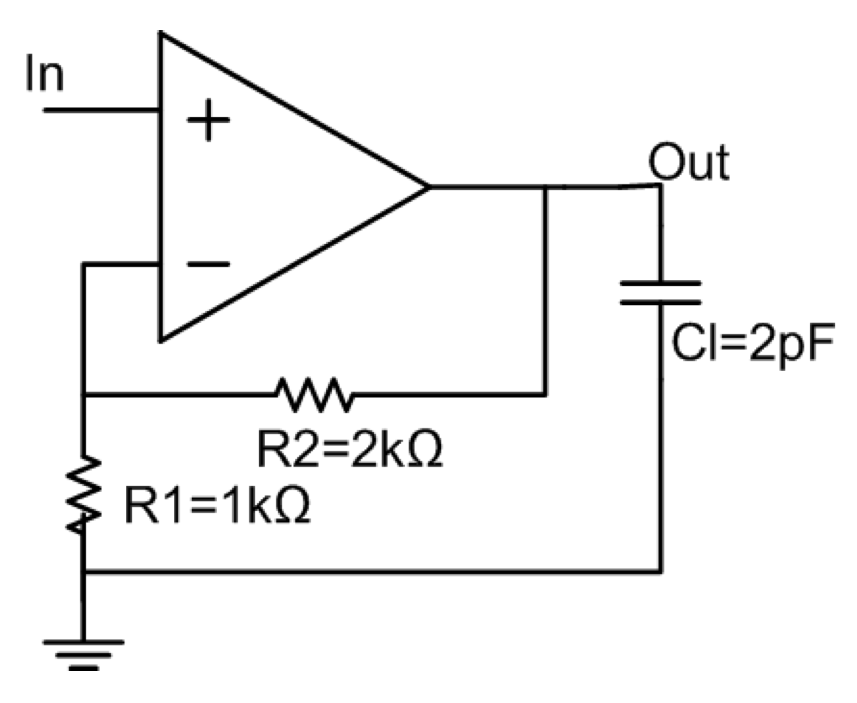}}
\caption{Simplified internal schematic of the opamp and feedback
    configuration used for data generation.  Parasitics not shown.}
 \label{fig:opamp}
\end{figure}

Our first example is causal dynamic inversion of an op-amp using data from a detailed high-dimensional 
simulation model. Figure~\ref{fig:opamp} presents the schematic of the op-amp at the transistor level, and the device in the closed loop simulation used for identification.  
For the task of dynamic inversion,
we seek to recover $v_{\textrm{in}}$ from $v_{\textrm{out}}$.  To
accomplish this we identify a dynamical system with $v_{\textrm{out}}$
as its input and $v_{\textrm{in}}$ as its output.

The identification data-set consists of two trials.  Each trial is a
simulation in response to $v_{\textrm{in}}$ being the sum of five
sinusoids.  The frequencies were chosen randomly within the
bandwidth of the amplifier (60 MHz). The trials
have different biases and amplitudes, and excite saturation in
the amplifier.  

Model complexity is varied in terms of state and input dimension, and polynomial degree. Our input
and surrogate state signals are in ``autoregressive moving-average'' form, parameterized by a positive integer $n$:
$
  \tilde y(t) = v_{\textrm{in}}(t), \:
  \tilde u(t) = \begin{bmatrix} v_{\textrm{out}}(t) & \ldots &
    v_{\textrm{out}}(t+1-n)\end{bmatrix}', \: 
  \tilde x(t) = \begin{bmatrix} v_{\textrm{in}}(t) & \ldots &
    v_{\textrm{in}}(t+1-n)\end{bmatrix}'.\notag
$
We search over models with $f$ being affine in the input and
$e,f$ of varying degree $d$ in the state $x(t)$, with $g$ fixed $g(x(t),u(t)) = x_1(t)$.
For all such models, minimizing equation error $J_{EE}$ resulted in an unstable model. 
Table~\ref{tab:opamp} presents the results using RIE for varying model complexities. 
The minimizers of the local RIE dramatically improve in performance as
the model complexity increases.


\begin{table}
\centering
\caption{Performance ($J_{\textrm{perf}}$) for Op-Amp dynamic inversion using local RIE with models of varying order $n$ and degree $d$.} 
\label{tab:opamp}
{\small \begin{tabular}{|l|ccc|}
    \hline
    $n \setminus d$  & 1 &  3 & 5\\
    \hline
    1  & 86.47 & 94.162& 94.22\\
    2 & 23.82& 21.90& 21.69\\
    3 & 11.80& 8.31& 7.99\\
    \hline
  \end{tabular}} 
\end{table}

\subsection{Identification of a Pneumatic Piston from Experiment}
Our second example is the identification of a double-acting
pneumatic cylinder under conditions simulating heavy load.  


Each chamber is separated from a control valve by approximately 25 cm
of tubing.  A compressor supplies high pressure (between 0.6 and 0.8
MPa) which is then regulated to a 0.5 MPa supply pressure to the valves, which have internal pressure control mechanisms. Both ends of the piston are fixed to a bracket. The dynamics of the system
include changes in supply pressure due to flow,
filling the chambers, the internal dynamics of the
valves and elastic bending of the rail to which the piston and bracket
are fixed.  The pressure is measured at the inlet of both chambers
(denoted $p_1(t)$ and $p_2(t)$ resp.).  A load cell between the
bracket and piston measures the force exerted on the bracket
($f_o(t)$).  We denote the commanded pressure signals as $v_1(t)$ and
$v_2(t)$ for valve 1 and 2 resp.
Finally, the supply pressure is measured near the inlet of the valves
($p_s(t)$). The input to the system is a scalar signal $f_d(t)$ which represents the
desired force in Newtons.

The experiment consisted of the application of seven chirp signals
of varying amplitudes being applied to the system.  These chirps
signals swept linearly in instantaneous frequency from approximately 3 to 40 Hz and peak desired force in the range 20 to 100 N.
The identification procedures were trained on data sets with
$20,60,100$ and $130$ N being the peak absolute desired force.
These models were then validated on trials with $40, 80$ and $120$ N
being the peak forces desired.
The experiments were performed at a sample-rate of 1 kHz.  
The desired
and realized force were taken to be the input and output signals and the internal pressures and
output force were chosen as the surrogate states:
$
\tilde u(t) = \begin{bmatrix} f_d(t) & f_d(t-1) \end{bmatrix}', \quad \tilde y(t) = f_o(t),  \tilde x(t)
= \begin{bmatrix}p_1(t) & p_2(t) & p_s(t) & f_o(t)\end{bmatrix}'.
$
We fit several discrete time models of the form
\eqref{mi5}, \eqref{mi6}.  We examine polynomial models of increasing
complexity, beginning with a linear model then  input
affine models with various degrees in $x$.  In all cases the
output map $g(\cdot,\cdot)$ is fixed to $g(x(t),u(t)) = x_4(t)$.

\begin{figure}
  \subfigure[Data and Simulation.]{\includegraphics[width=0.5\textwidth]{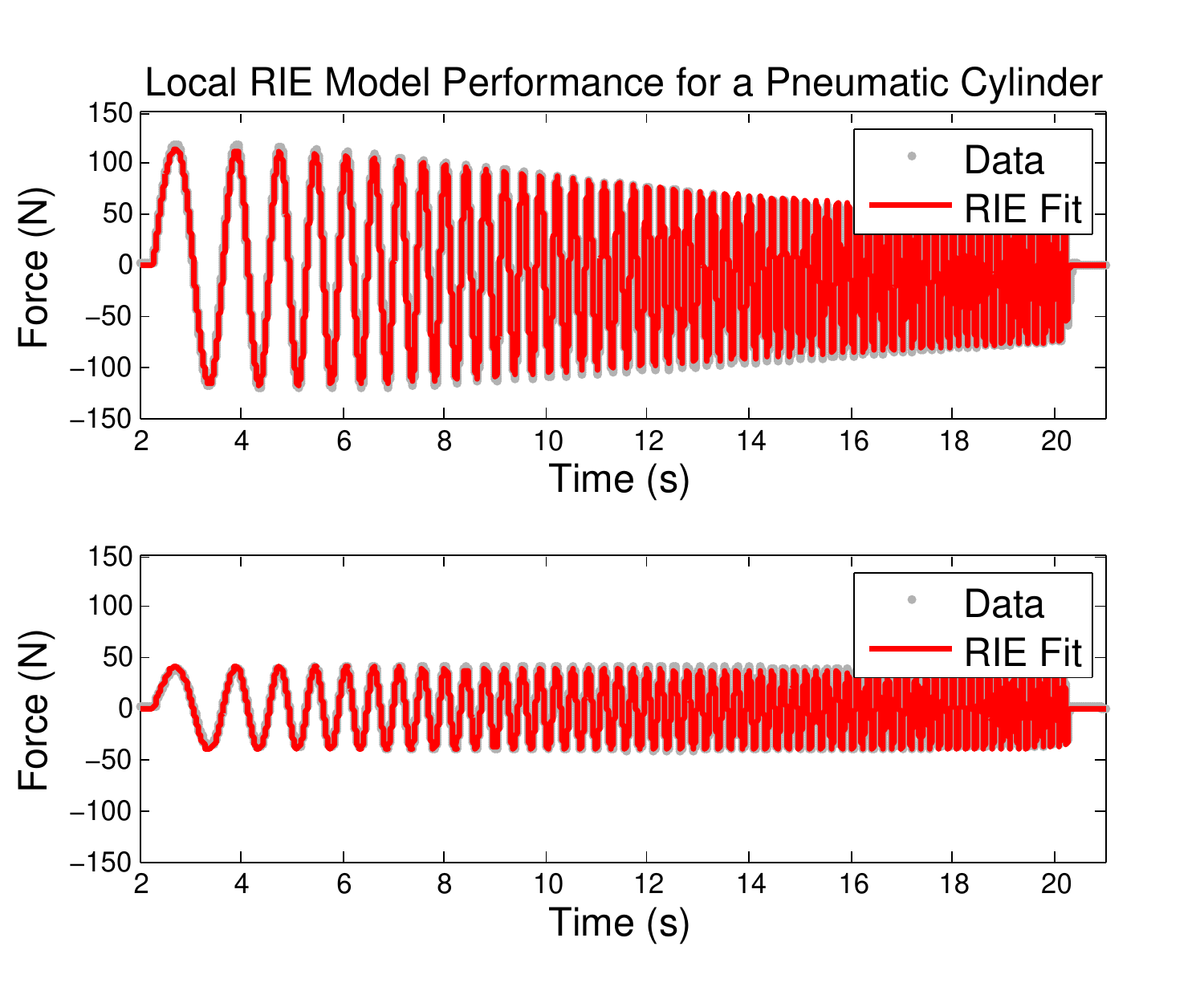}}
  \caption{Model performance using the local RIE: simulation
    (solid-red) compared against observed output (gray dots) for two
    different input amplitudes from nearly identical initial conditions.  The response does not simply scale,
    indicating nonlinearity.}
  \label{fig:pneumatic-fit}
\end{figure}

The computational complexity of minimizing the RIE scales with the number of data-points for nonlinear
fits. As a result, we sub-select 10,000 data-points for optimizing the
local RIE.  To ensure a fair comparison in these cases, we fit
an equation error model on both the same 10,000 data-points and the
whole data-set and report the better of the two scores.
Table~\ref{tab:pneumatic} compares the results, in order of increasing
model complexity. Figure~\ref{fig:pneumatic-fit} compares the observed $\tilde y(t)$ to
the open loop simulation of the best performing model from the local RIE.  Note that the
two plots are responses to differently scaled versions of the same
input, so nonlinearity is clearly apparent.

%
%
%
  \begin{table}
    \centering
\caption{Comparison of performance ($J_{\textrm{perf}}$) for models of the
  pneumatic piston on validation data. \label{tab:pneumatic}}
    \begin{tabular}{|l|c|c|}
    \hline
    Model Class & Equation Error & Local RIE\\
    \hline
    Linear & 29.31 & 24.62 \\
    Polynomial (deg$_x$($e$)=1, deg$_x$($f$)=1) & 14.34 &  22.07 \\
        Polynomial (deg$_x$($e$)=3, deg$_x$($f$)=1) & 81.26 &  14.84\\
    Polynomial (deg$_x$($e$)=3, deg$_x$($f$)=3)  & Divergent & 6.64 \\
    \hline
  \end{tabular}
\end{table}


We observe that the local RIE under-performs equation error for the first
polynomial model structure. However, with increasing model
complexity the equation error minimization at first improves, then rapidly degrades
in performance and becomes unstable. On the other hand, the performance of the local RIE minimization steadily
improves with increasing model complexity. These observations are consistent with the op-amp dynamic inversion.

One interpretation of this performance is that the stability constraints and the ``robust'' fidelity bounds act as a form of regularization, and mitigate the usual harmful effects of over-fitting and reduce the need for structure selection or ``pruning'' of regressors in autoregressive nonlinear models \cite{billings1982identification}. Indeed, for linear systems we have observed that these constraints confer a bias towards ``more stable'' models \cite{manchester2012stable}, \cite{Umenberger16}. Quantifying this effect will be a topic of future research.

%% file: cdc2010MIack.tex
We would like to thank Ryuma Niiyama for assistance with the pneumatic cylinder experiments, and Zhipeng Li, Yan Li and Vladimir Marko Stojanovic for their collaboration in providing the RF amplifier data set.

%% file: TMM_technote.bbl
\begin{thebibliography}{10}
\providecommand{\url}[1]{#1}
\csname url@samestyle\endcsname
\providecommand{\newblock}{\relax}
\providecommand{\bibinfo}[2]{#2}
\providecommand{\BIBentrySTDinterwordspacing}{\spaceskip=0pt\relax}
\providecommand{\BIBentryALTinterwordstretchfactor}{4}
\providecommand{\BIBentryALTinterwordspacing}{\spaceskip=\fontdimen2\font plus
\BIBentryALTinterwordstretchfactor\fontdimen3\font minus
  \fontdimen4\font\relax}
\providecommand{\BIBforeignlanguage}[2]{{%
\expandafter\ifx\csname l@#1\endcsname\relax
\typeout{** WARNING: IEEEtran.bst: No hyphenation pattern has been}%
\typeout{** loaded for the language `#1'. Using the pattern for}%
\typeout{** the default language instead.}%
\else
\language=\csname l@#1\endcsname
\fi
#2}}
\providecommand{\BIBdecl}{\relax}
\BIBdecl

\bibitem{antoulas2005approximation}
A.~C. Antoulas, \emph{Approximation of large-scale dynamical systems}.\hskip
  1em plus 0.5em minus 0.4em\relax Society for Industrial and Applied
  Mathematics, 2005, vol.~6.

\bibitem{Holmes98}
P.~Holmes, J.~Lumley, and G.~Berkooz, \emph{Turbulence, Coherent Structures,
  Dynamical Systems and Symmetry}.\hskip 1em plus 0.5em minus 0.4em\relax
  Cambridge University Press, 1998.

\bibitem{Lall02}
S.~Lall, J.~E. Marsden, and S.~Glavaski, ``A subspace approach to balanced
  truncation for model reduction of nonlinear control systems,''
  \emph{International Journal of Robust and Nonlinear Control}, vol.~12, no.~6,
  pp. 519--535, 2002.

\bibitem{lucia2004reduced}
D.~J. Lucia, P.~S. Beran, and W.~A. Silva, ``Reduced-order modeling: new
  approaches for computational physics,'' \emph{Progress in Aerospace
  Sciences}, vol.~40, no.~1, pp. 51--117, 2004.

\bibitem{Sjoberg95}
J.~Sj\"oberg, Q.~Zhang, L.~Ljung, A.~Benveniste, B.~Delyon, P.-Y. Glorennec,
  H.~Hjalmarsson, and A.~Juditsky, ``Nonlinear black-box modeling in system
  identification: a unified overview,'' \emph{Automatica}, vol.~31, no.~12, pp.
  1691--1724, 1995.

\bibitem{Ljung99}
L.~Ljung, \emph{System Identification: Theory for the User}, 3rd~ed.\hskip 1em
  plus 0.5em minus 0.4em\relax Englewood Cliffs, New Jersey, USA: Prentice
  Hall, 1999.

\bibitem{pearson2006nonlinear}
R.~K. Pearson, ``Nonlinear empirical modeling techniques,'' \emph{Computers \&
  chemical engineering}, vol.~30, no.~10, pp. 1514--1528, 2006.

\bibitem{Ljung10}
L.~Ljung, ``Perspectives on system identification,'' \emph{Annual Reviews in
  Control}, vol.~34, no.~1, pp. 1 -- 12, 2010.

\bibitem{Schon10}
T.~Sch{\" o}n, A.~Wills, and B.~Ninness, ``System identification of nonlinear
  state-space models,'' \emph{Automatica}, vol.~47, no.~1, pp. 39--49, 2010.

\bibitem{billings1982identification}
S.~Billings and S.~Fakhouri, ``Identification of systems containing linear
  dynamic and static nonlinear elements,'' \emph{Automatica}, vol.~18, no.~1,
  pp. 15--26, 1982.

\bibitem{billings2013nonlinear}
S.~Billings, \emph{Nonlinear System Identification: NARMAX Methods in the Time,
  Frequency, and Spatio-Temporal Domains}.\hskip 1em plus 0.5em minus
  0.4em\relax Wiley, 2013.

\bibitem{Doyle02}
F.~J. Doyle, R.~K. Pearson, and B.~A. Ogunnaike, \emph{Identification and
  control using {V}olterra models}.\hskip 1em plus 0.5em minus 0.4em\relax
  Springer Verlag, 2002.

\bibitem{Regalia95}
P.~A. Regalia and P.~Stoica, ``Stability of multivariable least-squares
  models,'' \emph{IEEE Signal Processing Letters}, vol.~2, no.~10, pp.
  195--196, 1995.

\bibitem{maciejowski1995guaranteed}
J.~M. Maciejowski, ``Guaranteed stability with subspace methods,''
  \emph{Systems \& Control Letters}, vol.~26, no.~2, pp. 153--156, 1995.

\bibitem{van2001identification}
T.~Van~Gestel, J.~A. Suykens, P.~Van~Dooren, and B.~De~Moor, ``Identification
  of stable models in subspace identification by using regularization,''
  \emph{IEEE Transactions on Automatic Control}, vol.~46, no.~9, pp.
  1416--1420, 2001.

\bibitem{Lacy03}
S.~L. Lacy and D.~S. Bernstein, ``Subspace identification with guaranteed
  stability using constrained optimization,'' \emph{IEEE Transactions on
  Automatic Control}, vol.~48, no.~7, pp. 1259--1263, 2003.

\bibitem{besselink2013model}
B.~Besselink, N.~van~de Wouw, and H.~Nijmeijer, ``Model reduction for nonlinear
  systems with incremental gain or passivity properties,'' \emph{Automatica},
  vol.~49, no.~4, pp. 861--872, 2013.

\bibitem{soderstrom1975uniqueness}
T.~S{\"o}derstr{\"o}m, ``On the uniqueness of maximum likelihood
  identification,'' \emph{Automatica}, vol.~11, no.~2, pp. 193--197, 1975.

\bibitem{Megretski08}
A.~Megretski, ``Convex optimization in robust identification of nonlinear
  feedback,'' in \emph{Proceedings of the 47th IEEE Conference on Decision and
  Control}, Cancun, Mexico, Dec 9-11 2008, pp. 1370--1374.

\bibitem{Bond10}
B.~Bond, Z.~Mahmood, Y.~Li, R.~Sredojevic, A.~Megretski, V.~Stojanovic,
  Y.~Avniel, and L.~Daniel, ``Compact modeling of nonlinear analog circuits
  using system identification via semidefinite programming and incremental
  stability certification,'' \emph{IEEE Transactions on Computer-Aided Design
  of Integrated Circuits and Systems}, vol.~29, no.~8, pp. 1149 --1162, aug.
  2010.

\bibitem{tobenkin2010convex}
M.~M. Tobenkin, I.~R. Manchester, J.~Wang, A.~Megretski, and R.~Tedrake,
  ``Convex optimization in identification of stable non-linear state space
  models,'' in \emph{IEEE Conference on Decision and Control (CDC)}, 2010, pp.
  7232--7237.

\bibitem{manchester2012stable}
I.~R. Manchester, M.~M. Tobenkin, and A.~Megretski, ``Stable nonlinear system
  identification: Convexity, model class, and consistency,'' in \emph{IFAC
  Symposium on System Identification}, vol.~16, no.~1, 2012, pp. 328--333.

\bibitem{tobenkin2013stable}
M.~M. Tobenkin, I.~R. Manchester, and A.~Megretski, ``Stable nonlinear
  identification from noisy repeated experiments via convex optimization,'' in
  \emph{American Control Conference (ACC)}, 2013, pp. 3936--3941.

\bibitem{tobenkin2014robustness}
M.~M. Tobenkin, ``Robustness analysis for identification and control of
  nonlinear systems,'' Ph.D. dissertation, Massachusetts Institute of
  Technology, 2014.

\bibitem{polik2007survey}
I.~P{\'o}lik and T.~Terlaky, ``A survey of the {S}-lemma,'' \emph{SIAM review},
  vol.~49, no.~3, pp. 371--418, 2007.

\bibitem{lemarechal2001lagrangian}
C.~Lemar{\'e}chal, ``Lagrangian relaxation,'' in \emph{Computational
  Combinatorial Optimization}.\hskip 1em plus 0.5em minus 0.4em\relax Springer,
  2001, pp. 112--156.

\bibitem{boyd2009convex}
S.~Boyd and L.~Vandenberghe, \emph{Convex optimization}.\hskip 1em plus 0.5em
  minus 0.4em\relax Cambridge university press, 2009.

\bibitem{poznyak2010advanced}
A.~Poznyak, \emph{Advanced Mathematical Tools for Control Engineers: Volume 1:
  Deterministic Systems}.\hskip 1em plus 0.5em minus 0.4em\relax Elsevier,
  2010.

\bibitem{van1996subspace}
P.~Van~Overschee and B.~De~Moor, \emph{Subspace identification for linear
  systems: theory, implementation, applications}.\hskip 1em plus 0.5em minus
  0.4em\relax Kluwer academic publishers, 1996.

\bibitem{Lohmiller98}
W.~Lohmiller and J.-J.~E. Slotine, ``On contraction analysis for non-linear
  systems,'' \emph{Automatica}, vol.~34, no.~6, pp. 683--696, June 1998.

\bibitem{parrilo2003semidefinite}
P.~A. Parrilo, ``Semidefinite programming relaxations for semialgebraic
  problems,'' \emph{Mathematical programming}, vol.~96, pp. 293--320, 2003.

\bibitem{Yalmip}
J.~Lofberg, ``Yalmip: A toolbox for modeling and optimization in matlab,'' in
  \emph{Computer Aided Control Systems Design, 2004 IEEE International
  Symposium on}.\hskip 1em plus 0.5em minus 0.4em\relax IEEE, 2004, pp.
  284--289.

\bibitem{sturm1999using}
J.~F. Sturm, ``Using {SeDuMi} 1.02, a matlab toolbox for optimization over
  symmetric cones,'' \emph{Optimization methods and software}, vol.~11, no.
  1-4, pp. 625--653, 1999.

\bibitem{Umenberger16}
J.~Umenberger and I.~R. Manchester, ``Specialized algorithm for identification
  of stable linear systems using {L}agrangian relaxation,'' in \emph{Proc. 2016
  American Control Conference}, Boston, MA, 2016.

\end{thebibliography}
